\documentclass[sigconf]{acmart}

\usepackage{booktabs} 

\setcopyright{rightsretained}

\acmDOI{10.475/123_4}

\acmISBN{123-4567-24-567/08/06}

\acmConference[CoNEXT'17]{ACM CoNEXT'17 Student Workshop}{December 2017}{Seoul/Incheon, South Korea}
\acmYear{2017}
\copyrightyear{2017}

\acmPrice{15.00}

\begin{document}
\title{Enabling Ultra-Low Delay Teleorchestras using Software Defined Networking}

\author{Emmanouil Lakiotakis}
\affiliation{%
  \institution{FORTH-ICS}
  \institution{University of Crete}
  \country{Greece}
}
\email{manoslak@ics.forth.gr}

\author{Christos Liaskos}
\affiliation{%
  \institution{FORTH-ICS}
  \country{Greece}
}
\email{cliaskos@ics.forth.gr}

\author{Xenofontas Dimitropoulos}
\affiliation{%
  \institution{FORTH-ICS}
  \institution{University of Crete}
  \country{Greece}
}
\email{fontas@ics.forth.gr}

\begin{abstract}
Ultra-low delay sensitive applications can afford delay only at the level of msec. An example of this application class are the Networked Music Performance (NMP) systems that describe a live music performance by geographically separate musicians over the Internet. The present work proposes a novel architecture for NMP systems, where the key-innovation is the close collaboration between the network and the application. Using SDN principles, the applications are enabled to adapt their internal audio signal processing, in order to cope with network delay increase. Thus, affordable end-to-end delay is provided to NMP users, even under considerable network congestion.
\end{abstract}

%
%



\maketitle

\section{Introduction}

\label{section:intro}
Networked Music Performance (NMP) systems, describe the process where musicians located in different places perform synchronized via the Internet~\cite{lazzaro2001case}. NMPs belong to ultra-low delay sensitive applications due to the latency requirements they have. In NMP services the maximum affordable delay between the transmitted and the finally reproduced signal should be up to 25 ms. This constraint is denoted as the Ensemble Performance Threshold (EPT)~\cite{schuett2002effects}.

There are many factors that affect end-to-end delay in NMP systems, which are grouped in two categories: network delay and audio processing delay. Network delay expresses the delay due to data transmission via the network, attributed to physical propagation and the network operating state. On the other hand, the audio processing delay is introduced by the audio capturing, processing and encoding methods, for each transmitter/receiver pair. Due to the delay that audio encoders yield (about 100 ms), they are avoided in NMP services; instead, raw audio is preferred~\cite{akoumianakis_musinet_2014,goto1910virtual}.

In this work, our goal is to endow NMP systems with resistance against traffic congestion and link failure cases. To make this feasible, our system introduces collaboration between the NMP application and the network. In more detail, the proposed architecture exploits the flexibility for dynamic network reconfiguration and global view of the network condition that Software Defined Networking (SDN) offers in order to achieve the acceptable latency for NMP~\cite{durner2015performance,gorlatch_improving_2014,kobayashi_maturing_2014}. The SDN controller, that orchestrates the network, is responsible for the data path setup that will carry the audio between each transmitter/receiver pair. Additionally, the controller is responsible for rerouting audio flows in case of congested paths. In cases where the whole network is congested, offering no feasible paths for NMP, the SDN controller communicates with the NMP application, at both the transmitter and receiver sides, in order to switch to another audio configuration set, thus decreasing the audio processing delay and compensating the increased network latency. Raw audio is used, instead of  encoders~\cite{maribondo_avoiding_2016,wabnik2009error,valin2016high}, due to their latency constraints. The proposed architecture, avoids network bandwidth waste, since each node receives a single version of the initial signal, instead of different encoded versions of the same signal~\cite{akoumianakis_musinet_2014,alexandraki2008towards,baltas_ultra_2014,carot2007network,carot2006network,carot2007networked,
xiaoyuan_gu_network_centric_2005}. The proposed Software-Defined NMP solution can be also applied in more generic use cases encountered in the real Internet, e.g., considering the domains (Autonomous Systems) between the sender and the receiver as ``big switches'', abstracted with the use of classic tunneling mechanisms~\cite{kotronis2016stitching,kotronis2014control}. This approach is scalable since it requires only a number of points of presence anywhere in the network where paths can be programmatically set up and monitored either in the form of overlay tunnels or physical links.

Similar approaches try to guarantee QoS via applying certain policies on the traffic queues of the switches/routers~\cite{kumar_user_2013,sharma_implementing_2014,durner_performance_2015,sieber_network_2015,egilmez_distributed_2014,adami_network_2015}, based on different criteria such as source and destination IP address, transport protocol or generally applying  either Type of Service (TOS) matching~\cite{sieber_network_2015,egilmez_distributed_2014,adami_network_2015,sharma_demonstrating_2014} or DiffServ approach~\cite{tomovic_sdn_2014}. On the other hand, we apply smart routing in the network based on real-time measurements instead of incorporating metrics that reflect the difference between the transmitted and the required by the user audio quality ~\cite{mu2016scalable,koumaras_-service_2016} combined with the adjustment of application-tunable parameters, in order to deal with sudden increases in network delay. We evaluated our system in an emulation environment, demonstrating the advantages of the proposed architecture.
\section{The SDN Teleorchestra System} \label{sdn_teleorchestra}
As described in Section~\ref{section:intro}, end to end delay in NMPs is the summary of network and audio processing delay. In mathematical form, end to end delay in NMPs for similar user audio equipment can be modeled as:
\begin{equation}
d_{end-to-end}=2\times d_{sound-card} + d_{n}
\label{eq:1}
\end{equation}
where $d_{end-to-end}$ represents end-to-end delay, $d_{sound-card}$ shows the delay due to sound-card in transmitter and receiver and $d_{n}$ expresses the network delay. Delay due to audio processing is alternatively called as blocking delay. Equation (\ref{eq:2}) describes the blocking delay evaluation:
\begin{equation}
\label{eq:2}
d_{blocking-delay}= \frac{frame\ size}{sampling\ rate} + d_{0}
\end{equation}
In equation (\ref{eq:2}), frame size denotes the size of audio packets that a user sound-card can process per hardware clock tick, and the sampling rate represents the number of samples the sound-card acquires per second. Finally, $d_{0}$ is a constant delay that is due to the sound-card's hardware quality. For blocking delay decrease, the fraction between frame size and sampling rate should be also decreased.

\begin{figure}[!t]
\centering
\includegraphics[width=0.6\columnwidth]{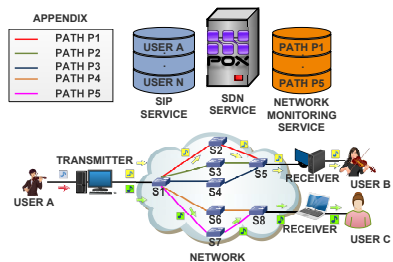}
\caption{The proposed system architecture.}
\label{fig:1}
\end{figure}

The proposed architecture consists of three components: a SIP, an SDN and a Network Monitoring service as shown in Fig.~\ref{fig:1}. The SIP service initially measures the audio performance of each user for different audio settings and stores this information. Additionally, it classifies each user as premium or regular based on the privileges that he has. Premium users do not accept low quality transmissions compared with regular users. This category contains musicians that require high quality audio or users (audience) that may have paid for low latency according to SLAs. SIP triggers the application when audio modification is required \cite{nam_towards_2014,nurmela2007session,ali2013session,sinnreich_internet_2006,camarillo_evaluation_2003}.

The SDN service describes the controller functionality that installs paths for audio transmission. The number of paths towards each user and the path characteristics are defined by the user classification. Each path consists of OVS switches that are instructed by SDN service via the OpenFlow protocol~\cite{akyildiz_roadmap_2014,mckeown_openflow:_2008}. Also, the SDN service reroutes audio flows when a better path is available. Finally, the Network Monitoring service measures network delay for all paths between users by periodic UDP packet transmission between them.
When a new user participates in our system, the SIP service creates an audio profile of the user for various audio parameter combinations. Moreover, it classifies the user as either premium or regular as discussed. Based on the classification results, the SDN service assigns a single path from the audio source to the user and audio transmission starts, and keeps a tunable number of backup paths at the ready. The Network Monitoring service continuously monitors all paths and the SDN service reroutes audio flows to a path that surpasses the active one in performance. If all paths are congested resulting in over-EPT end to end delay, the SIP service contacts the application side in order to switch to an audio mode that has less blocking delay. When the network recovers from congestion, the application returns to the previous audio configuration.

\section{System Evaluation} \label{system_evaluation}
\begin{figure}[!t]
\centering
\includegraphics[width=0.9\columnwidth]{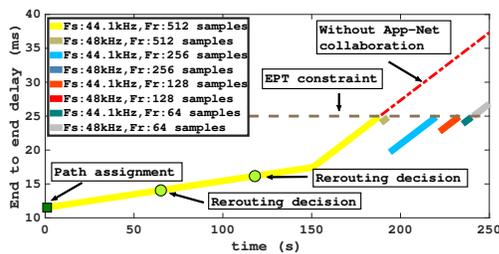}
\caption{End-to-end delay evaluation.}
\label{fig:2}
\end{figure}

To evaluate our system, we used real live audio, and Mininet ~\cite{lantz_network_2010,sharma2014mininet} to emulate a realistic network setup. The topology is shown in Fig.~\ref{fig:1}. POX ~\cite{ligia_rodrigues_prete_simulation_2014,kaur2014network,bagewadi2014towards,sukhveer_network_????,silvan_pox_????} is selected as the SDN controller. We used Netem traffic control tool in order to increase delay in each path. To avoid redundant reroutes, we used a threshold of at least 2 ms lower latency as the criterion for taking rerouting decisions among available paths. The results of our experimental evaluation are depicted in Fig.~\ref{fig:2}. $F_s$ denotes sampling rate and $F_r$ the frame size. The proposed method can also be applied in conventional Internet topologies enabling guaranteed end-to-end delay via inter-domain routing ~\cite{Liaskos1,Liaskos2,Liaskos3,Liaskos4}.

Assume User B declares interest for receiving audio from User A. There are three paths between User A and User B in the test topology. At the beginning, the SDN service assigns path $P_1$ for audio transmission with $F_s$ equal to $44.1$ kHz and $F_r$ set to $512$ samples. As we increase the network delay in the path, the SDN service reroutes audio flows to paths $P_2$ (at $t=65$ s) and $P_3$ (at $t=118$ s).  At $t=189$ s, the SIP service informs the application to switch to an audio mode with lower blocking delay ($F_s$ equal to $48$ KHz and unaltered $F_r$). Assuming that the network delay increases further, and in order to keep end-to-end delay below EPT, at $t=194$ s, the application modifies $F_r$ to $256$ samples via interaction with the SIP service. This process takes place during the experimental setup for various audio modes. When the network delay is very high, our system results in best effort delay (after $t=241$ s), given the available audio modes and network condition. Comparing our method with the case that application and network could not interact, we achieve delay improvement by $29.6\%$ in end-to-end delay.
\begin{acks}
This work has been funded by the European Research Council Grant Agreement no. 338402.
\end{acks}

\bibliographystyle{ACM-Reference-Format}

\end{document}